\documentclass[a4paper,11pt]{article}
\pdfoutput=1 

\usepackage{jcappub} 

\usepackage{float}
\usepackage[T1]{fontenc} 
\usepackage{graphicx}

\usepackage{tikz}
\usetikzlibrary{arrows.meta, calc, positioning}
\hypersetup{colorlinks=true,urlcolor=blue,citecolor=blue,linkcolor=blue,menucolor=blue,anchorcolor=blue,filecolor=blue}

\title{\boldmath Cosmological implications of \texorpdfstring{$f(T, B)$}{} gravity: Constraints from recent observations}

\author[a]{Laxmipriya Pati,}
\author[b,1]{Santosh V. Lohakare,\note{Corresponding author}}
\author[c,d]{S. K. Maurya} 

\affiliation[a]{Institute of Physics, University of Tartu, W. Ostwaldi 1, 50411 Tartu, Estonia}
\affiliation[b]{Department of Mathematics, School of Advanced Sciences, Vellore Institute of Technology, Vellore, 632014, Tamil Nadu, India}
\affiliation[c]{Department of Mathematical and Physical Sciences, College of Arts and Sciences, University of Nizwa, P.O. Box 33, Nizwa 616, Sultanate of Oman}
\affiliation[d]{Research Center of Astrophysics and Cosmology, Khazar University, Baku, AZ1096, 41 Mehseti Street, Azerbaijan}

\emailAdd{lpriyapati1995@gmail.com}
\emailAdd{lohakare.santosh@vit.ac.in}
\emailAdd{sunil@unizwa.edu.om}

\abstract{In this work, we examine the theoretical framework of modified teleparallel gravity with the inclusion of the boundary term in the action and investigate its cosmological implications by considering the power-law model $f(T, B) = -T + \alpha (-B)^\beta$, with the aim of addressing late-time accelerated expansion and the dark energy. In this context, $T$ denotes the torsion scalar and $B$ represents the boundary term, whose presence allows for departures from standard teleparallel dynamics and provides a unified description that connects torsion and curvature-based formulations, reproducing $f(T)$ and $f(R)$ gravity in appropriate limits. The viability of the model is assessed by confronting its theoretical predictions with observational data while constraining the cosmological and model parameters through a Markov Chain Monte Carlo (MCMC) using cosmic chronometers (CC), the Pantheon plus sample (PPS), and the DESI baryon acoustic oscillation (BAO) Data Release 2 (DR2) datasets, and comparing its performance with the standard $\Lambda$CDM model. Our results, for the obtained constraints $\beta$ from the model parameter, strongly indicate that, within the adopted parameterization and prior ranges, the observational data favor a non-zero correction to the boundary term. The Akaike information criterion (AIC) analysis shows that the combined CC+PPS dataset mildly favors the $f(T, B)$ model, suggesting an improved phenomenological fit to late-time observations over $\Lambda$CDM, while in contrast, the Bayesian information criterion (BIC) favors the $\Lambda$CDM, as the $f(T, B)$ model is penalized due to an additional free parameter introduced through the initial conditions required to solve the cosmological equations, aiming to constrain from statistical analysis to compare with the observations. Further, our model predicts that the Hubble constant lies between the Planck and SH0ES determinations, which is again consistent with current late-time observations. In addition, the background cosmological quantities suggest that the model exhibits a dynamical phantom-divide crossing while remaining consistent with late-time observations and yields a viable expansion history and characteristic dark energy evolution.}

\begin{document}
\maketitle
\flushbottom

\section{Introduction} \label{SEC I}

    Despite the success of general relativity (GR), several theoretical and observational challenges continue to motivate the cosmological community to explore extensions of the standard cosmological model \cite{Perivolaropoulos_2022_95, Sotiriou_2010_82}. On the theoretical side, GR does not provide a quantum description of gravity or a complete ultraviolet description of the theory, nor does it explain the microscopic origin of dark matter or dark energy (DE). On the observational side, the discrepancy between early-Universe determinations of $H_0$ inferred within the $\Lambda$CDM model and late-Universe, distance-ladder measurements remains an open problem \cite{DiValentino_2021_38} and see Ref.\ \cite{Cai_2026_26} for a recent Hubble tension review. In addition, the $S_{8}$ tension \cite{DiValentino_2020_131, Cosmoverse_white_paper_2025} and the cosmological constant problem \cite{Weinberg_1989_61} further motivate the exploration of extensions to the standard cosmological framework \cite{Abdalla_2022_34}. Although these discrepancies do not prove that GR fails on cosmological scales, they make it natural to ask how modified gravity may affect both the background evolution and cosmological observables \cite{Clifton_2012_513, Saridakis_2021_2105.12582, Nojiri_2017_692, Nojiri_2011_505}.

    In the standard formulation of GR, gravity is described by the curvature of spacetime associated with the Levi-Civita connection, which is torsion-free and metric-compatible \cite{Misner_1973_book}. An equivalent description can instead be constructed in the teleparallel framework, where curvature and nonmetricity vanish, and gravity is encoded in the torsion scalar $T$; this formulation is known as the teleparallel equivalent of general relativity (TEGR) \cite{Aldrovandi_2013_TG_INTRO}. A third dynamically equivalent formulation is provided by symmetric teleparallel gravity, where curvature and torsion vanish, and the gravitational dynamics are encoded in the nonmetricity scalar $(Q)$. This framework is known as the symmetric teleparallel equivalent of general relativity (STEGR)~\cite{Nester_1999}. These three formulations constitute different geometric representations of the same gravitational dynamics at the level of GR, with their corresponding actions differing only by boundary terms and known as the ``geometric trinity of gravity'' \cite{Jimenez-2019}. However, their nonlinear extensions, such as a nonlinear function of torsion scalar or nonmetricity scalar, are in general, no longer dynamically equivalent and can lead to distinct cosmological evolutions and observational predictions \cite{Bahamonde_2015_92, Heisenberg_2023_1066_Review}. In particular, while $f(T)$ and $f(Q)$ gravity represent genuine modifications of TEGR and STEGR, respectively, the inclusion of the appropriate boundary terms allows one to recover curvature-based $f(R)$ gravity for specific combinations. In the teleparallel case, since $R=-T+B$, the choice $f(-T+B)$ is equivalent to $f(R)$ gravity, whereas a generic $f(T,B)$ theory is not \cite{Bahamonde_2015_92, Kadam_2022_37_FTB}. Similarly, in the symmetric teleparallel case, the corresponding $f(Q,B)$ theory reduces to $f(R)$ gravity only for the specific combination $f(Q-B)$, while a general $f(Q,B)$ model represents a distinct class of modified gravity \cite{Capozziello_2023_83}.

    \begin{figure}[H]
        \begin{center}
        \begin{tikzpicture}[node distance=1cm, 
        every node/.style={font=\normalsize}, 
        roundnode/.style={circle, draw=black, minimum size=0.5cm, align=center},
        arrow/.style={-{Latex[length=3mm]}, thick}]

        \node[roundnode] (A) at (90:3cm) {$f(T,B)$};
        \node[roundnode] (B) at (0:3cm) {$f(T)$};
        \node[roundnode] (C) at (270:3cm) {GR};
        \node[roundnode] (D) at (180:3cm) {$f(R)$};

        \draw[arrow] (A) -- (B) node[midway, above, sloped] {$f = f(T)$};
        \draw[arrow] (B) -- (C) node[midway, below, sloped] {$f = T$};
        \draw[arrow] (D) -- (C) node[midway, below, sloped] {$f = R$};
        \draw[arrow] (A) -- (D) node[midway, above, sloped] {$f = f(-T + B)$};
        \end{tikzpicture}
        \caption{Comparison of modified gravity theories with GR, highlighting key differences and shared features in their predictions for gravitational phenomena.}
        \end{center}
        \label{FIG: intro}
    \end{figure}
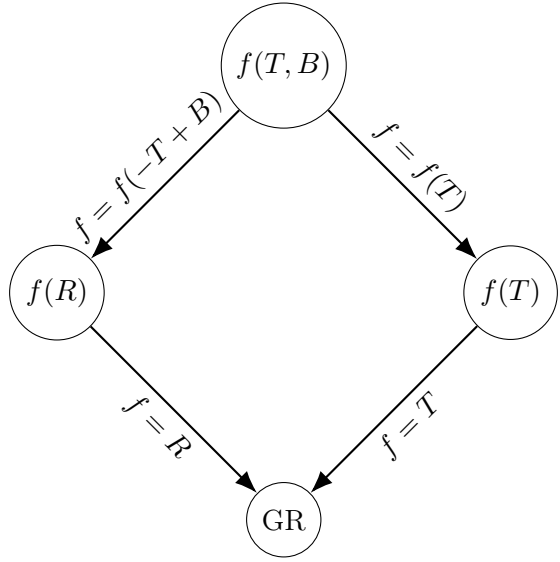

    Figure~\ref{FIG: intro} illustrates the hierarchical structure of modified theories of gravity within the $f(T, B)$ framework. It shows how $f(T, B)$ gravity provides a unifying extension that connects torsion-based and curvature-based formulations, while also allowing for genuinely new gravitational dynamics beyond GR. Unlike GR, which propagates two tensor polarizations of the graviton, nonlinear extensions introduce additional degrees of freedom. In particular, $f(T)$ or the equivalent scalar torsion theory is known to propagate extra modes beyond those of GR, although the precise number may depend on the background spacetime and the formulation adopted \cite{Li_2011_07, Ferraro_2018_97, Blagojevic_2020_102}. Likewise, previous studies indicate that $f(Q)$ or the equivalent scalar nonmetricity theory possesses a richer dynamical structure than GR. In particular, Ref.\ \cite{Chen_2026_113} reported seven propagating degrees of freedom in $f(Q)$; this result is in tension with other analyses \cite{Hu_2022_106, Tomonari_2024_84} and, therefore, warrants further independent confirmation. A cosmological perturbation study in Ref.\ \cite{Bahamonde_2021_81} indicates that generic $f(T, B)$ gravity possesses additional propagating modes beyond the two tensorial degrees of freedom of GR, potentially associated with local Lorentz-symmetry breaking, although a full Hamiltonian analysis is still required to establish the precise number of propagating degrees of freedom.

    A growing body of literature has explored whether modified gravity can alleviate cosmic tensions without violating physical assumptions, as reviewed in this paragraph. A viable $f(Q)$ model was tested against recent observational data, showing that nonmetricity-based modified gravity can provide a competitive alternative to $\Lambda$CDM at both the background and perturbation levels \cite{Anagnostopoulos_2021_822}. Ref.\ \cite{Boiza_2025_12} claimed that the coincident $f(Q)$ gravity can partially alleviate the $H_0$ and $S_{8}$ tensions by considering some well-motivated nonmetricity functional via performing a Bayesian analysis, while at the same time the authors \cite{Yang_2024_533} found evidence that non-coincident gauge formulations provide a better fit to cosmological observations than $\Lambda$CDM. A perturbative extension of $f(Q)$ gravity that modifies structure-growth dynamics while leaving the background cosmological evolution unchanged, showing that it can alleviate the $\sigma_{8}$ tension \cite{Li_2026_07}. Ref.\ \cite{Ren_2022_932} found how torsional gravity addresses the Hubble tension by employing Gaussian-process reconstruction and the effective field theory formalism of teleparallel gravity to derive observationally favored $f(T)$ models. Ref.\ \cite{Banerjee_2023_108} studied cosmological applications of ghost-free gravitational scalar-tensor theories and showed that their late-time deviations from $\Lambda$CDM can increase the inferred value of $H_{0}$, offering a viable mechanism to alleviate the Hubble tension. Ref.\ \cite{Bouhmadi-Lopez_2026_12} constrained $f(T)$ models using recent conservation data, showing that late-time torsional modifications can partially alleviate the $H_0$ tension.

    Although the tetrad-based formulation of $f(T,B)$ gravity is technically more involved due to the presence of fourth-order field equations, recent years have witnessed growing interest in exploring its cosmological implications. The authors in Ref.\ \cite{Koivisto_2019_5} analyzed the particle spectrum by deriving the corresponding propagating degrees of freedom in the most general quadratic torsion-based teleparallel gravity. Ref.\ \cite{Bahamonde_2021_81} addressed the theoretical and observational viability condition in $f(T, B)$ gravity, derived the tensor, vector, and scalar perturbation equations, and the effective gravitational coupling to assess via cosmological perturbation. The propagating degrees of freedom in the most general quadratic torsion-based teleparallel gravity and theoretical implications are discussed in \cite{Koivisto_2019_5, Bahamonde_2015_92}. Within this framework, exact cosmological solutions, including reconstruction methods and thermodynamic properties, have been investigated in \cite{Bahamonde-2016-19} and the cosmological dynamics in a scalar field non-minimally coupled to teleparallel theory explored in \cite{Bahamonde_2015_92}. Cosmological solutions in $f(T, B)$ gravity have also been obtained using the Noether symmetry approach \cite{Bahamonde-2017-77-2}. In addition, the asymptotic behavior and stability of cosmological solutions have been studied through dynamical-systems analysis for a spatially flat Friedmann--Lema\^{i}tre--Robertson--Walker (FLRW) background in \cite{Paliathanasis_2021_136}.  
    
    Despite investigations of the theoretical and dynamical aspects of $f(T,B)$ gravity, its viability in light of recent late-time observational data remains comparatively less explored. In particular, a combined analysis using cosmic chronometer (CC) measurements, the Pantheon plus sample (PPS), and the recent DESI BAO DR2 data provides a useful opportunity to test whether a simple power-law boundary-term correction can reproduce the observed expansion history and distinguish itself from $\Lambda$CDM through diagnostic tools such as the deceleration parameter, the effective equation of state (EoS), statefinder parameters, and the ${\rm Om}(z)$ diagnostic. One of the main goals of the present study is to start with a well-motivated $f(T, B)$ model, derive the relevant cosmological observables, and confront the model with current observational data to examine whether it can alleviate or possibly resolve existing cosmological tensions. This work addresses this gap by deriving the relevant cosmological equations for the model $f(T, B)=-T+\alpha (-B)^\beta$, constraining its parameters using late-time datasets, and assessing its phenomenological performance relative to $\Lambda$CDM. We consider a spatially flat FLRW background, within which a generic $f(T, B)$ model leads to cosmological field equations containing higher-order derivatives of the scale factor, thereby allowing the framework to reconstruct or effectively mimic different cosmological expansion histories.

    The article is organized as follows. Section~\ref{sec2:Geometric formalism of teleparallel gravity} introduces the basic geometric preliminaries of teleparallel gravity. Section~\ref{sec3:Metric teleparallel action with the inclusion of a boundary term} presents the modified action and the corresponding field equations. The background cosmological equations are discussed in Section~\ref{sec4:Cosmology in $f(T,B)$ gravity}. In Section~\ref{sec5:Cosmological Observations with Numerical Approach}, we perform the observational analysis using different datasets and examine the associated cosmological observables. Finally, Section~\ref{sec6:Conclusion} summarizes the main findings and discusses their implications.

\section{Geometric formalism of teleparallel gravity} \label{sec2:Geometric formalism of teleparallel gravity}
    In teleparallel gravity, the spacetime metric is constructed from the tetrad $e^{a}{}_{\mu}$, its inverse $E_{a}{}^{\mu}$, and the Minkowski metric with signature $\eta^{ab}={\rm diag}(-1,+1,+1,+1)$ is
        \begin{align}
        \label{Lowerindex_Spacetime_Metric}
            g_{\mu\nu} &= e^{a}{}_{\mu} e^{b}{}_{\nu} \eta_{ab} \,\,,
        \end{align}
    while its inverse is given by
        \begin{align}
        \label{Upperindex_Spacetime_Metric}
            g^{\mu\nu} &= E_{a}{}^{\mu} E_{b}{}^{\nu} \eta^{ab} \, ,  
        \end{align}
    where $E_{a}{}^{\mu}$ denotes the inverse tetrad.

    The tetrad and cotetrad satisfy the orthonormal relations
        \begin{align}
            E_{a}{}^{\mu} e^{b}{}_{\mu} &= \delta^{b}_{a} \,,\label{deltanm} \\
            E_{a}{}^{\mu} e^{a}{}_{\nu} &= \delta^{\mu}_{\nu} \,\label{deltamunu}.
        \end{align}
        
    Furthermore, the determinant of the metric is related to the  tetrad determinant as
        \begin{align}
            e=\sqrt{-g}. 
        \end{align}

    The fundamental geometrical object in metric teleparallel gravity is the torsion tensor, defined as the antisymmetric part of the teleparallel connection. In the present framework, we adopt the vanishing spin connection, under which the torsion tensor reduces to the form
        \begin{align}
            T^{a}{}_{\mu\nu} &= \Gamma^{a}{}_{\mu\nu}-\Gamma^{a}{}_{\nu\mu}=
            \partial_{\mu} e^{a}{}_{\nu} - \partial_{\nu}e^{a}{}_{\mu} \,.
        \end{align}
        
    A fully spacetime-indexed torsion tensor is obtained by projecting onto the coordinate basis using the inverse tetrad,
        \begin{align}
            \label{Torsion_Tensor}
            T^{\lambda}{}_{\mu\nu} = E_{a}{}^{\lambda} T^{a}{}_{\mu\nu} \,.
        \end{align}
        
    The torsion vector, corresponding to the trace of the torsion tensor, takes the form
        \begin{align}
            T_{\mu}=T^{\lambda}{}_{\lambda\mu} \,.
        \end{align}
        
    It is worth noting that the tetrad postulate establishes the relation between the teleparallel connection and the Levi--Civita connection,
    \begin{align}
        \Gamma_{\lambda}{}^{\mu}{}_{\rho} = {\mathring{\Gamma}}^{\mu}{}_{\lambda\rho} + K_{\lambda}{}^{\mu}{}_{\rho}\,,
    \end{align}
    where $K_{\lambda}{}^{\mu}{}_{\rho}$ denotes the contortion tensor, defined as
        \begin{align}
	       K_{\mu}{}^{\lambda}{}_{\nu}&=\frac{1}{2}\left(T^{\lambda}{}_{\mu\nu}-T_{\nu\mu}{}^{\lambda}+T_{\mu}{}^{\lambda}{}_{\nu}\right).
        \end{align}

    The Ricci scalar $R$ constructed from the Levi--Civita connection can be expressed in terms of the tetrad field as
        \begin{align}
            R(e) = -\left(\frac{1}{4}T^{abc}T_{abc}+\frac{1}{2}T^{abc}T_{bac}-T^aT_a\right) + 
            \frac{2}{e}\partial_\mu (e T^\mu) \,,
        \end{align}
    which can equivalently be rewritten as
        \begin{align}
            R(e) = - S^{abc}T_{abc} + \frac{2}{e}\partial_\mu (e T^\mu) \,,
            \label{ricciS}
        \end{align}
    where the contraction of the superpotential tensor $S^{abc}$ with the torsion tensor defines the torsion scalar,
        \begin{align}
            \label{Torsion_Scalar}
            T=S^{abc}T_{abc} \,,
        \end{align}
    with the superpotential tensor given explicitly by
        \begin{align}
            S^{abc} = \frac{1}{4}(T^{abc}-T^{bac}-T^{cab})+\frac{1}{2}(\eta^{ac}T^b-\eta^{ab}T^c) \,,
        \end{align}
    which in spacetime can be written as
        \begin{align}
            2S_{\sigma}{}^{\mu\nu} = K_{\sigma}{}^{\mu\nu} - \delta^{\mu}_{\sigma}T^{\nu} + 
            \delta^{\nu}_{\sigma}T^{\mu} \,.
            \label{S}
        \end{align}
        
    Consequently, equation~\eqref{ricciS} can be written in the compact form
        \begin{align}
            R(e) = - T + \frac{2}{e}\partial_\mu (e T^\mu) \,,
            \label{ricciT}
        \end{align}
    which provides the fundamental relation underlying teleparallel gravity.

    Although $R$ is invariant under local Lorentz transformations, neither $T$ nor $B$ term individually shares this invariance. However, their specific combination remains invariant. For convenience, we define the boundary term as
        \begin{align}
            \label{Boundary term}
            B = \frac{2}{e}\partial_\mu (e T^\mu) = 2 \nabla_\mu T^\mu \,.
        \end{align}

\section{Metric teleparallel action with the inclusion of a boundary term} \label{sec3:Metric teleparallel action with the inclusion of a boundary term}

    The action functional, by including a boundary term \eqref{Boundary term} that consists of a second-order derivative of the fundamental variable tetrad inside the generic function, can be written as
        \begin{align}
            \label{f(T,B) action}
            S_{\rm TB} = \int 
            \left[ 
            \frac{1}{\kappa^2}f(T,B) + L_{\rm m}
            \right] e\, d^4x \,,
         \end{align}
    where $\kappa^2 = 16 \pi G$, $G$ being the Newtonian constant, $f(T,B)$ is an arbitrary function, and $L_{\rm m}$ denotes the matter Lagrangian density. The above action generalizes both teleparallel gravity and curvature-based modified gravity theories. In particular, by neglecting the term $B$, one recovers the $f(T)$ gravity. Moreover, using the relation $R = -T + B$, an appropriate choice of the function $f(T,B)$ allows one to reproduce $f(R)$ gravity, thereby establishing the equivalence between torsion and curvature-based descriptions at the action level.
         
    The corresponding energy-momentum tensor is defined via the variational derivative of the matter action with respect to the tetrad field,
        \begin{align}
            \Theta_{a}{}^{\lambda}=\frac{1}{e} \frac{\delta (eL_m)}{\delta e^{a}{}_{\lambda}} \,.
        \end{align}

    Varying the action \eqref{f(T,B) action} with respect to the tetrad field $e^{a}{}_{\mu}$ yields the tensorial field equations
    \begin{multline}
    \label{General tensor equations}
        2eE_{a}{}^{\lambda}\Box f_{B}-2eE_{a}{}^{\sigma}\nabla^{\lambda}\nabla_{\sigma}f_{B}+
        eBf_{B}E_{a}{}^{\lambda} + 4e\Big[(\partial_{\mu}f_{B})+(\partial_{\mu}f_{T})\Big]S_{a}{}^{\mu\lambda} 
        \\
        +4\partial_{\mu}(e S_{a}{}^{\mu\lambda})f_{T}-4ef_{T}T^{\sigma}{}_{\mu a}S_{\sigma}{}^{\lambda\mu}-
        e f E_{a}{}^{\lambda} = 16\pi e \Theta_{a}{}^{\lambda}.
    \end{multline}
    
    Contracting the tetrad field equation~\eqref{General tensor equations} with a tetrad $e^{a}{}_{\nu}$, one obtains the purely spacetime-indexed field equations as,
    \begin{multline}
    \label{fieldeq}
            2e\delta_{\nu}^{\lambda}\Box f_{B}-2e\nabla^{\lambda}\nabla_{\nu}f_{B}+
          e B f_{B}\delta_{\nu}^{\lambda} + 
          4e\Big[(\partial_{\mu}f_{B})+(\partial_{\mu}f_{T})\Big]S_{\nu}{}^{\mu\lambda}
          \\
          +4e^{a}{}_{\nu}\partial_{\mu}(e S_{a}{}^{\mu\lambda})f_{T} - 
          4 e f_{T}T^{\sigma}{}_{\mu \nu}S_{\sigma}{}^{\lambda\mu} - 
          e f \delta_{\nu}^{\lambda} = 16\pi e \Theta_{\nu}{}^{\lambda} \,,
    \end{multline}
    where $f_T\equiv \partial f/\partial T$ and $f_B\equiv \partial f/\partial B$ denote the partial derivatives of $f(T,B)$ with respect to $T$ and $B$, respectively. Moreover, $\Theta_{\nu}{}^{\lambda}=e^{a}{}_{\nu}\Theta_{a}{}^{\lambda}$ represents the standard energy-momentum tensor.

    When studying a modified gravity theory, an essential first step is to examine its theoretical consistency and observational viability. It is known that the $f(T, B)$ gravity was motivated by $f(R)$ and can be viewed as its teleparallel counterpart, hence it shares several important features with the latter. Like $f(R)$ theory, in addition to the two standard massless tensor modes of GR, $f(T, B)$ gravity propagates an extra scalar degree of freedom. The mass of this scalar mode is approximately proportional to $\sim \frac{1}{ \sqrt{-f_{BB}}}$. Consequently, the conditions
    \begin{align}
        f_{T} < 0, \quad \text{and} \quad f_{BB}<0
    \end{align}
    in the field Eq.\ \eqref{fieldeq} are required to ensure a positive effective gravitational coupling and to avoid ghost and tachyonic instabilities, respectively \cite{Koivisto_2019_5}.

    In the following, we consider limiting cases at the level of the field equations that reproduce $f(T)$ and $f(R)$ gravity, respectively. In particular, the teleparallel equivalent of $f(R)$ gravity can be obtained by restricting the functional dependence to the specific combination
        \begin{equation}
            \tilde{f}(T,B) := \tilde{f}(-T + B) = \tilde{f}(R) \,.
        \end{equation}
        
     In this limit, the torsional and boundary contributions combine to reproduce curvature-based dynamics, thereby establishing the equivalence between the teleparallel formulation and $f(R)$ gravity. Consequently, the higher-order contributions associated with $R$ arise through $B$, while the pure torsion sector encodes the complementary dynamics.

\section{Cosmology in \texorpdfstring{$f(T,B)$}{} gravity} \label{sec4:Cosmology in $f(T,B)$ gravity}

    We consider a spatially flat FLRW background spacetime described by the metric
        \begin{equation}
            \label{FLRW_Metric}
            ds^2=-dt^2+a(t)^2(dx^2+dy^2+dz^2)\,,
        \end{equation}
    where $a(t)$ is the scale factor characterizing the expansion history of the Universe. The corresponding tetrad field consistent with the spacetime metric relations \eqref{Lowerindex_Spacetime_Metric} and \eqref{Upperindex_Spacetime_Metric} can be chosen as
        \begin{equation}
        \label{FLRW_tetrad}
            e^{a}{}_{\mu}=\mbox{diag}(1,a(t),a(t),a(t))\,.
        \end{equation}
        
    For the cosmological background \eqref{FLRW_Metric} and \eqref{FLRW_tetrad}, the torsion scalar defined in equation~\eqref{Torsion_Scalar} reduces to
        \begin{equation}\label{torsionscalar_frw}
            T = 6H^2\,,
        \end{equation}
    while the associated boundary term defined in \eqref{Boundary term} reduces to
        \begin{equation}\label{boundaryscalar_frw}
            B = 6\left(3H^2+\dot{H}\right)\,.
        \end{equation}
        
    Consequently, the combination of the above $T$ and $B$ reproduces the $R$,
        \begin{equation}
            R = -T+B = 6\left(\dot{H} + 2H^2\right)\,.
        \end{equation}
          
    Substituting the tetrad \eqref{FLRW_tetrad}, together with the torsion scalar \eqref{torsionscalar_frw}, and boundary term \eqref{boundaryscalar_frw}, into the general field equations \eqref{General tensor equations}, one obtains the modified cosmological field equations,
    \begin{subequations}
    \label{Modified_Friedmann_Equation}
        \begin{align}
            -3H^2\left(3f_B + 2f_T\right) + 3H\dot{f}_B - 3\dot{H} f_B + \frac{1}{2}f &= \kappa^2\rho_{\rm m}\label{Friedmann_1}\,,\\
            -\left(3H^2+\dot{H}\right)\left(3f_B + 2f_T\right) - 2H\dot{f}_T + \ddot{f}_B + \frac{1}{2}f &= -\kappa^2 p_{\rm m}\label{Friedmann_2}\,.
        \end{align}
    \end{subequations}

    Without loss of generality, one can perform the redefinition
        \begin{equation}
            f(T,B) \rightarrow -T+\tilde{f}(T, B),
        \end{equation}
    in the action \eqref{f(T,B) action}, which isolates the TEGR and ensures manifest diffeomorphism invariance of the modified gravitational sector. For convenience, these equations can be recast into the standard Friedmann form,
    \begin{subequations}
        \begin{eqnarray}
            3H^2 &=& \kappa^2 \left(\rho_{\rm m}+\rho_{\mathrm{DE}}\right) = \kappa^2 \rho_{\text{eff}}\,, \label{eq: first_field}\\
            3H^2+2\dot{H} &=& -\kappa^2\left(p_{\rm m}+p_{\mathrm{DE}}\right)= -\kappa^2 p_{\text{eff}}\,,
        \end{eqnarray}
    \end{subequations}
    where $\rho_{\mathrm{DE}}$ represents the effective energy density 
        \begin{eqnarray}
            \kappa^2 \rho_{\mathrm{DE}} &=& 3H^2(3\tilde{f}_B+2\tilde{f}_T)-3H \dot{\tilde{f}}_B+3\dot{H}\tilde{f}_B-\frac{1}{2}\tilde{f} \,, \label{eq:friedmann_mod}
             \end{eqnarray}
    and $p_{\mathrm{DE}}$ represent DE pressure
        \begin{eqnarray}
            \kappa^2 p_{\mathrm{DE}} &=& -(3H^2+\dot{H})(3\tilde{f}_B+2\tilde{f}_T)-2H\dot{\tilde{f}}_T+\ddot{\tilde{f}}_B +\frac{1}{2}\tilde{f} \,,
        \end{eqnarray}
    from the modified gravity sector. 

    By combining the modified Friedmann equations \eqref{Modified_Friedmann_Equation}, one obtains the dynamical relation
        \begin{equation}
            2\dot{H}=-\kappa^2\left(\rho_{\rm m} + p_{\rm m} + \rho_{\mathrm{DE}} + p_{\mathrm{DE}}\right)\,.
        \end{equation}

    The DE sector satisfies the conservation equation
        \begin{equation}
            \dot{\rho}_{\mathrm{DE}}+3H\left(\rho_{\mathrm{DE}}+p_{\mathrm{DE}}\right) = 0\,,
        \end{equation}
    which allows the definition of a DE EoS parameter,
        \begin{eqnarray}
            \omega_{\mathrm{DE}} &=& \frac{p_{\mathrm{DE}}}{\rho_{\mathrm{DE}}} \nonumber\\ 
            &=& -1+\frac{\ddot{\tilde{f}}_B-3H\dot{\tilde{f}}_B-2\dot{H}\tilde{f}_T-2H\dot{\tilde{f}}_T}{3H^2\left(3\tilde{f}_B+2\tilde{f}_T\right)-3H\dot{\tilde{f}}_B+3\dot{H}\tilde{f}_B-\frac{1}{2}\tilde{f}}\, \, . \label{EoS_func}
        \end{eqnarray}

\section{Cosmological datasets and numerical analysis} \label{sec5:Cosmological Observations with Numerical Approach}
        We investigate the $f(T, B)$ cosmological model through a comprehensive analysis of three complementary late-time datasets, selected for their robustness and independence. These datasets include: (1) CC, which deliver model-independent measurements of the Hubble parameter $H(z)$ derived from the differential age evolution of elliptical galaxies \cite{Moresco_2022_25}; (2) PPS, encompassing 1701 Type Ia supernovae (SNe Ia) spanning redshifts up to $z \sim 2.3$ \cite{Brout_2022_938, Scolnic_2022_938}; and (3) the DESI BAO Data Release 2 (DR2), providing precise distance measurements from galaxy clustering and Lyman alpha forest observations across the redshift range $0.3 \leq z \leq 2.33$~\cite{Adame_2024_DESI_collaboration, DESI_DR2}. In this work, we make use of several independent cosmological observations, namely CC, PPS and DESI BAO data, to constrain the model parameters. Combining these datasets not only improves the overall statistical power of the analysis but also helps minimize possible systematic biases without requiring a fiducial cosmological model~\cite{Agostino_2018_98, Agostino_2019_99, Lohakare_2023_40_CQG}. The following sections describe the datasets employed in this study and the corresponding likelihoods adopted in the parameter estimation. We also discuss the DESI DR2 anisotropic clustering data, whose precision represents a significant improvement over previous BAO measurements. The novelty of this work lies in testing a simple power-law $f(T,B)$ model against a combined set of recent late-time observations, including DESI BAO DR2, and in using kinematic, statefinder, and ${\rm Om}(z)$ diagnostics to assess its deviation from $\Lambda$CDM.

\subsection{Cosmological datasets}
\subsubsection{Cosmic chronometers (CC)} \label{SEC III a}
    The CC method provides a direct way to estimate the Hubble parameter $H(z)$ from the differential age evolution of passively evolving galaxies. In this approach, massive early-type galaxies observed within small redshift intervals are used as cosmic clocks, so that the expansion rate can be inferred from the relation between redshift and cosmic time, rather than from an assumed cosmological background \cite{Jimenez_2002_573}. Its main advantage is that it gives observational constraints on $H(z)$ without requiring the adoption of a specific cosmological model.

    In the present analysis, we use the latest CC compilation \cite{Moresco_2022_25, Lohakare_2023_40_CQG}, together with the associated covariance matrix. This is important because the CC measurements are affected not only by statistical errors but also by systematic uncertainties in stellar population synthesis modelling. These include, for example, uncertainties in the initial mass function, stellar libraries, metallicity, and other assumptions that enter into the age determination of old stellar populations. Following developments in earlier CC studies \cite{Moresco_2012_2012_006, Moresco_2015_450, Moresco_2016_2016_014}, the use of the covariance matrix allows these correlated uncertainties to be included consistently. The resulting $H(z)$ data therefore provide a useful and relatively model-independent input for testing the background expansion history of the model.

\subsubsection{Type Ia supernovae (SNe Ia)} \label{SEC III b}
    Type Ia Supernovae (SNe Ia) are highly luminous stellar explosions, making them detectable across cosmological distances, even at redshifts approaching $z \approx 2.3$. Their role as one of the primary tools for measuring cosmological distances is largely due to the uniform nature of their spectral and photometric evolution, combined with the fact that they can be observed over a large fraction of the sky. The distance modulus ($\mu$) for SNe Ia is empirically derived from observational data using the relationship
        \begin{eqnarray}
            \mu = m_b - M\, ,
        \end{eqnarray}
    where $m_b$ represents the apparent magnitude, which serves as a normalization for the overall flux, and $M$ denotes the absolute magnitude. In the present analysis, $M$ is treated as a nuisance parameter and constrained simultaneously with the cosmological parameters. Theoretically, the distance modulus can also be expressed in terms of the luminosity distance $d_L(z)$ as
        \begin{eqnarray}
            \mu = 25 + 5 \log_{10}(d_L(z)) \, ,
        \end{eqnarray}
    
    The luminosity distance itself is defined by
        \begin{eqnarray}
            d_L(z) = (1 + z) \int_0^z \frac{c}{H(z')} \, dz' \, ,        
        \end{eqnarray}
    where $c$ is the speed of light in vacuum.

    Since the PPS dataset is calibrated using Cepheid observations from the SH0ES collaboration, an independent calibration of the supernova absolute magnitude is not required \cite{Riess_2022_934, Chantada_2023_107, Camarena_2020_2, Conley_2011_192}. This makes it possible to treat the $M$ as a free parameter in the analysis. Although recent work has highlighted potential issues with aspects of the calibration process \cite{Perivolaropoulos_2024_110, Freedman_2025_993}, the method remains widely used and continues to yield robust cosmological constraints.

\subsubsection{Baryon acoustic oscillation (BAO)} \label{SEC III c}
    The baryon acoustic oscillations (BAO) are formed due to the density fluctuations in the baryonic matter, which are created by acoustic waves. The BAO create a specific scale in the large-scale structure, thus acting as a cosmic yardstick. We utilize the DESI Data Release 2 (DR2) dataset, detailed in Ref.~\cite{DESI_DR2}, which provides precise measurements of the comoving distance at the drag epoch, expressed as $D_M / r_d$, and the Hubble distance, $D_H / r_d$, where $r_d$ represents the comoving sound horizon at baryon decoupling---the maximum distance sound waves could travel from the big bang until that epoch. In scenarios with lower signal-to-noise ratios, the volume-averaged distance $D_V / r_d$ is used. When analyzing DESI BAO data independently, only the combined parameter $r_d H_0$ can be constrained. However, integrating DESI DR2 measurements with complementary cosmological datasets enables independent constraints on $r_d$ and the Hubble constant $H_0$.

    The DESI DR2 BAO dataset~\cite{DESI_DR2} comprises two isotropic measurements, from the Bright Galaxy Survey (BGS) at $z_{\mathrm{eff}}=0.30$ and the quasar (QSO) sample at $z_{\mathrm{eff}}=1.49$, together with ten anisotropic BAO  measurements. The anisotropic constraints are obtained from the luminous red galaxy (LRG) samples at $z_{\mathrm{eff}}=0.51$ and $0.71$, the combined LRG+emission line galaxy (ELG) sample at $z_{\mathrm{eff}}=0.93$, the ELG sample at $z_{\mathrm{eff}}=1.32$, and the Lyman-$\alpha$ quasar sample at $z_{\mathrm{eff}}=2.33$. The expanded galaxy and quasar samples in DR2 lead to more precise BAO measurements than were available in earlier surveys. As a result, the data can constrain the transverse and radial distance scales, $D_M(z)$ and $D_H(z)$, independently, instead of only providing constraints on the combined quantity $D_V(z)$. Much of this gain comes from the improved signal-to-noise ratio of the DR2 quasar sample \cite{Adame_2024_DESI_collaboration}. 

    \begin{table*}[!htb]
    \centering
    \begin{tabular}{| c | c | c | c | c |}
        \hline
        Tracer & $z_{\mathrm{eff}}$ & $D_M/r_d$ & $D_H/r_d$ & $r \text{ or } D_V/r_d$ \\
        \hline
        BGS & 0.295 & - & - & $7.942 \pm 0.075$ \\
        LRG1 & 0.510 & $13.588 \pm 0.167$ & $21.863 \pm 0.425$ & $12.720 \pm 0.099$ \\
        LRG2 & 0.706 & $17.351 \pm 0.177$ & $19.455 \pm 0.330$ & $16.050 \pm 0.110$ \\
        LRG3+ELG1 & 0.934 & $21.576 \pm 0.152$ & $17.641 \pm 0.193$ & $19.721 \pm 0.091$ \\
        ELG2 & 1.321 & $27.601 \pm 0.318$ & $14.176 \pm 0.221$ & $24.252 \pm 0.174$ \\
        QSO & 1.484 & $30.512 \pm 0.760$ & $12.817 \pm 0.516$ & $26.055 \pm 0.398$ \\
        Lya & 2.330 & $38.988 \pm 0.531$ & $8.632 \pm 0.101$ & $31.267 \pm 0.256$ \\
        LRG3 & 0.922 & $21.648 \pm 0.178$ & $17.577 \pm 0.213$ & $19.656 \pm 0.105$ \\
        ELG1 & 0.955 & $21.707 \pm 0.335$ & $17.803 \pm 0.297$ & $20.008 \pm 0.183$ \\
        \hline
    \end{tabular}
        \caption{Statistical properties of the DESI DR2 BAO samples, detailing the effective redshift ($z_{\mathrm{eff}}$) for each measurement. The table includes corresponding distance ratios, presented as either the correlated pair $(D_M/r_d, D_H/r_d)$ with correlation coefficient $r$, or the spherically averaged distance $D_V/r_d$, offering a comprehensive overview of the cosmological constraints derived from these observations.}
    \label{Table:BAO stat DR2}
    \end{table*}

    The DESI DR2 observables, summarized in Table \ref{Table:BAO stat DR2}, are defined as follows:
    \begin{itemize}
        \item Hubble distance $D_H(z)$,
        \begin{equation}
            D_H(z) = \frac{c}{H(z)} \, .
        \end{equation}
        \item Angular diameter distance $D_A(z)$,
        \begin{equation}\
            D_A(z) = \frac{1}{(1+z)} \int_0^z \frac{c}{H(z')}dz' \, . 
        \end{equation}
        \item Transverse comoving distance $D_M(z)$,
        \begin{equation}
            D_M(z) = (1+z)D_A(z) \, .
        \end{equation}
        \item Volume averaged distance $D_V(z)$,
        \begin{equation}
            D_V(z) = \left[ (1+z)^2 D_A^2(z) \frac{c z}{H(z)} \right]^{1/3} \, .
        \end{equation}
    \end{itemize}
    These measurements enhance the precision of cosmological parameter estimation, leveraging the advanced data processing and fitting techniques of the DESI DR2 release.

\subsection{Exploring specific models in \texorpdfstring{$f(T, B)$ gravity }{}}
    In order to proceed further with our cosmological analysis, let us establish the form of the function in the $f(T, B)$ gravity. In phenomenological $f(T)$ gravity models, deviations from the TEGR are often introduced through power-law corrections involving $T$, although other functional forms are also possible. Here, we consider a minimal power-law correction to the TEGR sector in the $f(T,B)$ framework,
        \begin{equation}
            f(T, B) = -T + \alpha (-B)^\beta,
        \end{equation}
    where the constants $\alpha$ and $\beta$ describe the deviation from the TEGR limit by the amplitude and slope, respectively. In case of using dimensional variables, the constant $\alpha$ has proper units such that the action becomes dimensionally consistent. Even though this approach seems simple, this model exhibits rich cosmological dynamics, especially in the late Universe, in $f(T, B)$ gravity~\cite{Briffa_2023_39, Escamilla-Rivera_2019_37}. For $\alpha = 0$, the function becomes the TEGR theory that is identical to GR up to the redefinition of the gravitational constant. Related cosmological dynamics have previously been examined in the context of scalar-field DE models and, more recently, within teleparallel gravity frameworks~\cite{Copeland_1998_57, Ferreira_1997_79, Chen_2009_2009_04, Duchaniya_2025_85, Duchaniya_2023_83}.
    
    A critical aspect of this model lies in the behavior governed by the parameter $\beta$. Specifically,
    \begin{itemize}
        \item When $\beta = 0$, the model reduces to the standard $\Lambda$CDM cosmology, serving as a crucial consistency check and benchmark for comparison.
        \item If we take into account the case $\beta=1$, the modification is linear in $B$. Since $B$ represents a total derivative, this particular modification does not affect the background field equations under the constant $\alpha$ case, and the model reduces to GR/TEGR. Within the present observational analysis, the accelerating solutions are favoured in the region $\beta<1$, indicating a deviation from the linear boundary-term limit. This means that some deviation from the GR limit ($\beta = 1$) is needed \cite{Briffa_2023_39}.
    \end{itemize}

    Thus, we focus on scenarios where $\alpha \neq 0$ to investigate novel cosmological behaviors.

    To compute the theoretical Hubble parameter $H(z)$, we numerically solve the modified Friedmann equation. In a cosmological regime dominated by pressureless matter ($p_\mathrm{m} = 0$), the matter energy density evolves as
        \begin{equation}
            \rho_{\mathrm{m}}(z) = 3H_0^2 \Omega_{\mathrm{m}0} (1+z)^3,
        \end{equation}
    in which $z$ is the redshift of the cosmos, $a_0/a = 1+z$, ($a_0$ is the current scale factor, and $a$ is the scale factor at the time of emission of the observed light). Here $\Omega_{\mathrm{m}0}$ denotes the present-day matter density parameter, and we assume $\kappa^2=1$ throughout this work. Within this framework, the modified Friedmann equation for our model is formulated as
    \begin{eqnarray} \label{Eq: main ode}
            && -6 H^2 + 6 {H_0}^2 (z+1)^3 \Omega _{\text{m}0} 
            + \frac{\alpha \, 6^{\beta } (\beta -1) \left[ H \left((z+1) H' - 3 H\right)\right]^{\beta }}{\left((z+1) H' - 3 H\right)^2} \times \nonumber \\
            && \left[\left((z+1) \left[-\beta (z+1) H H'' - (\beta -1) (z+1) (H')^2 + (5 \beta -6) H H'\right] + 9 H^2\right)\right] = 0
    \end{eqnarray}

    In this framework, the equations are formulated in redshift space, with prime notation ($^\prime$) consistently denoting derivatives with respect to the cosmological redshift $z$, $({\rm i.e.}\,\,H^{'}(z)=\frac{dH}{dz})$. This transformation enables the examination of observational data in a cosmological framework that depends on redshift.

    The non-linear differential equation of order two for $H(z)$ has two boundary conditions, (i) the Hubble parameter at the present day $H(0) = H_0$, and (ii) the condition on the first-order derivative $H'(0) = \eta$. By allowing $H'(0)=\eta$ to vary independently, we explore the broader family of background solutions admitted by the $f(T,B)$ field equations, rather than restricting the model to solutions whose present-day slope matches that of $\Lambda$CDM. Here, the value of the parameter $\eta$ is not fixed a priori but is determined through an analysis based on Markov Chain Monte Carlo (MCMC). The parameter $\eta$ is responsible for setting the present-day value of the slope of the Hubble function, which in turn is equal to the present-day value of the deceleration parameter $q_0=-1+\eta/H_0$. At this point, a keen observer might ask why the second initial condition is an independent free cosmological parameter rather than being determined by the underlying dynamics. To answer this question, it is worth clarifying that the new additional free parameter introduced in the second initial condition is not an additional degree of freedom; rather, it is treated as a phenomenological initial parameter or it could be labels for different background solutions compatible with the same model. Since the field equations \eqref{Modified_Friedmann_Equation} do not uniquely determine $\eta$ without imposing additional assumptions on the cosmic evolution, we treat it as a phenomenological parameter and constrain it jointly with the model parameters using observational data.

    For the $f(T, B)$ gravity model under investigation, we adopt the following prior ranges for the model parameters in the MCMC analysis: $H_0 \in [50.0, 100.0] \, \mathrm{km} \, \mathrm{s}^{-1} \, \mathrm{Mpc}^{-1}$, $\Omega_{\mathrm{m}0} \in [0.0, 1.0]$, $\alpha \in [-5.0, 5.0]$, $\beta \in [-5.0, 5.0]$, $\eta \in [0.0, 50.0]$, and $M \in [-20.0, -18.0]$. These priors enable a robust comparison of the model against observational datasets, ensuring a comprehensive evaluation of its cosmological viability.    
    \begin{table} [H]
    \def\arraystretch{1.5}
        \centering
    \resizebox{1\textwidth}{!}{
    \begin{tabular}{|*{7}{c|}}\hline
        {\centering  \textbf{Datasets}} & $H_0$ & $\Omega_{\mathrm{m}0}$ & $\alpha$ & $\beta$ & $\eta$ & $M$ \\ [0.5ex]
    \hline \hline
        \parbox[c][0.8cm]{4cm}{\centering \textbf{CC+PPS}} & $70.132 \pm 1.151$ & $0.291^{+0.021}_{-0.023}$ & $1.312^{+0.107}_{-0.107}$ & $0.156^{+0.092}_{-0.092}$ & $31.188^{+0.328}_{-0.328}$ & 
        $-19.371^{+0.272}_{-0.272}$ \\
    \hline
        \parbox[c][0.8cm]{4cm}{\centering \textbf{CC+PPS+DESI DR2}} & $69.120^{+1.032}_{-1.032}$ & $0.320^{+0.018}_{-0.017}$ & $1.269^{+0.101}_{-0.101}$ & $0.162^{+0.078}_{-0.078}$ & $31.047^{+0.316}_{-0.316}$ & 
        $-19.252^{+0.135}_{-0.135}$ \\[0.5ex] 
    \hline
    \end{tabular}}
        \caption{Overview of cosmological parameters derived through a MCMC analysis using the CC, PPS, and DESI BAO datasets together. The table gives the posterior probability distribution for the Hubble parameter ($H_0$), present matter density parameter ($\Omega_{\mathrm{m}0}$), and the modified gravity parameters ($\alpha$, $\beta$, $\eta$, $M$) with $1\sigma$ error intervals.}
        \label{Table: parametrized values}
    \end{table}

    \begin{figure}[H]
    \centering
    \resizebox{\textwidth}{!}{%
        \includegraphics[width=170mm, height=210mm]{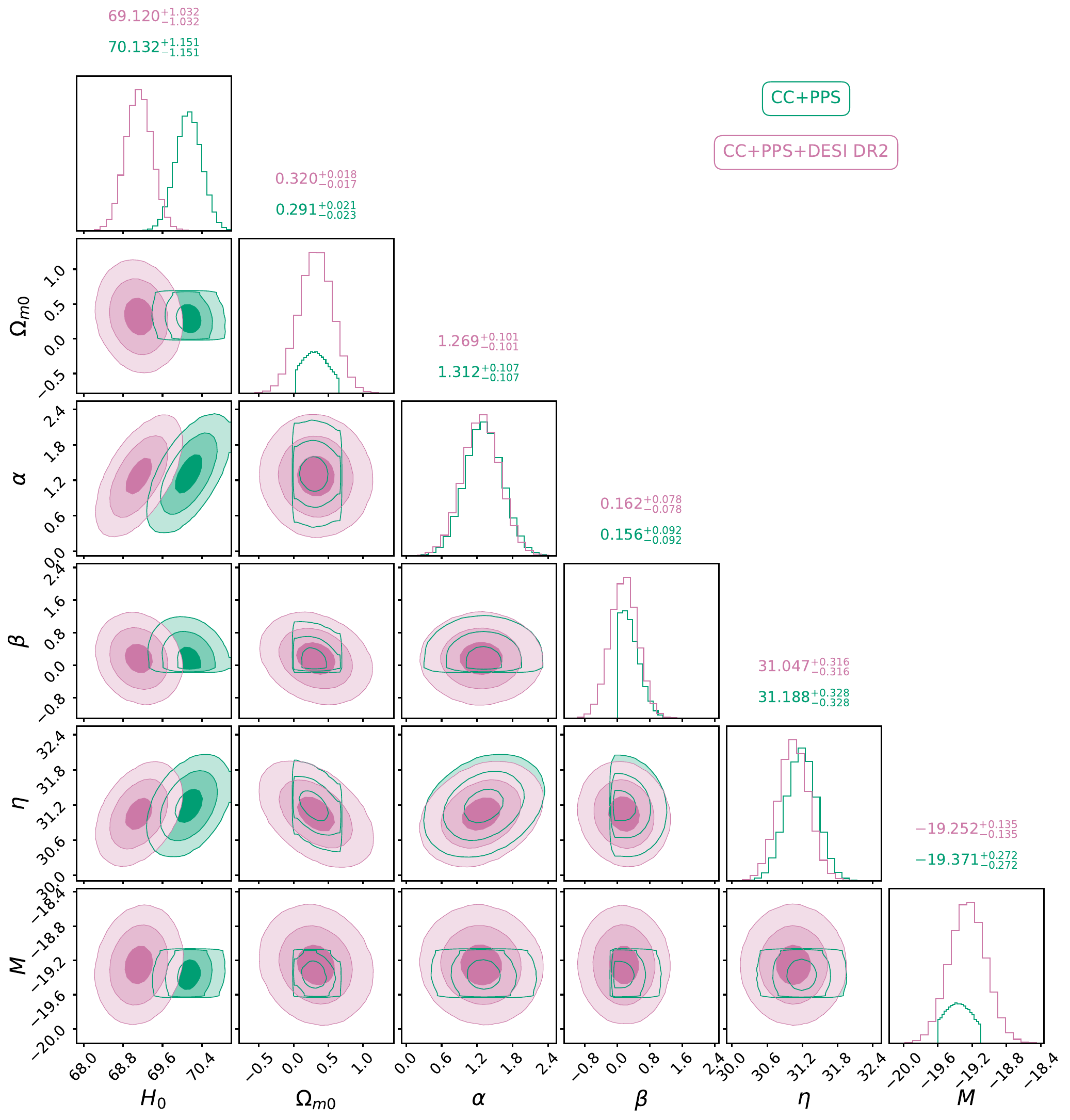}}
        \caption{Two-dimensional marginalized posterior probability distribution functions of model parameters. Here we show the 1$\sigma$ and 2$\sigma$ contours (representing $68\%$ and $95\%$ credible levels, respectively) of the main cosmological and model-dependent parameters Hubble constant ($H_0$), matter density parameter today ($\Omega_{\mathrm{m}0}$), and model-dependent parameters $\alpha$, $\beta$, $\eta$, and $M$. These results are obtained by using the CC+PPS and CC+PPS+DESI DR2 data sets.} 
    \label{FIG: MCMC_contour}
    \end{figure}

    Figure~\ref{FIG: MCMC_contour} shows the contour plot of the $1\sigma$ and $2\sigma$ confidence regions of CC+PPS and CC+PPS+DESI DR2, respectively. The MCMC results for the parameters that yield the best fit of the model are shown in Table \ref{Table: parametrized values}. The procedure used involves solving the Friedmann equations using two conditions to initialize them, in order to increase the reliability of the results. The above figure is entitled CC+PPS and uses data from CC and PPS, thus giving an indication of the constraints on the parameters involved. The $x$ and $y$-axes of the contour plot have been labeled using the major cosmological parameters $H_0$, $\Omega_{\text{m}0}$, $\alpha$, $\beta$, $\eta$ and $M$. Parameter ranges covered by the axes include a wide range of both negative and positive values. Figure~\ref{FIG: MCMC_contour} differs in color palettes and represents the interplay of the discussed parameters, where each section represents certain pairs of variables. Such a method of visualizing allows clearly distinguishing data clusters or categories and showing the statistical distribution. The shift from the green contours (CC+PPS) to the violet contours (CC+PPS+DESI DR2) reflects a decrease in the inferred Hubble constant, from $H_0=70.132\,\mathrm{km\,s^{-1}\,Mpc^{-1}}$ for CC+PPS to $H_0=69.120\,\mathrm{km\,s^{-1}\,Mpc^{-1}}$ for CC+PPS+DESI DR2. This indicates that including DESI DR2 distance measurements pulls the posterior constraints toward a lower and more conservative expansion rate, yielding an intermediate value between the Planck 2018 and local distance-ladder estimates. In this sense, we conclude that the Hubble constant prediction of our model lies between the Planck and SH0ES determinations, which is consistent with current late-time observations. The significant tightening of the purple contours in the $H_0 - \Omega_{\rm m0}$ plane shows how effectively the new BAO data help to overcome the geometric degeneracy, arising when only CC and PPS data are used. The inferred present matter density parameters are $\Omega_{\rm m0}=0.291^{+0.021}_{-0.023}$ and $\Omega_{\rm m0}=0.320\pm0.018$ for the CC+PPS and CC+PPS+DESI DR2 datasets, respectively. The inclusion of DESI DR2 results in a slightly higher matter density, indicating a slightly denser matter sector than the constraint from the CC+PPS combination alone. Moreover, the $1\sigma$ posterior intervals of the power-law index $\beta$ do not include zero for the considered datasets, with best-fit values $\beta\simeq 0.156$ and $\beta\simeq 0.162$ for CC+PPS and CC+PPS+DESI DR2, respectively. The obtained non-zero constraints $\beta$ for both data sets provide strong evidence that, within the adopted parameterization and prior ranges, the observational data favor a non-zero correction to the boundary term. The absolute magnitude of the supernovae ($M \simeq -19.371, -19.252$) is still stable and does not show any physical bias caused by the modified-gravity part of the framework.

    The MCMC simulation was performed numerically by solving the Friedmann equations under two different initial conditions. One initial assumption is the trivial one yet an acceptable one in cosmology that acts as a benchmark for our model. The other one was constrained from data using the MCMC technique. The use of two different initial assumptions not only confirms the self-consistency of the $f(T, B)$ model but also shows its flexibility in describing the accelerated expansion of the Universe. The contours derived from these two different initial assumptions provide interesting insights into parameter degeneracy and correlations in modified gravity models.

\subsection{Statistical evaluation of model performance}
    To assess the statistical performance of the $f(T, B)$ model in comparison to the standard $\Lambda$CDM framework, we employ a robust statistical framework utilizing the Akaike information criterion (AIC)~\cite{Akaike_1974_19}, the Bayesian information criterion (BIC)~\cite{Schwarz_1978_6}, and the minimum chi-squared statistic ($\chi^2_{\mathrm{min}}$). These metrics provide a balanced evaluation of model fit and complexity, ensuring a fair comparison by penalizing models with excessive parameters to mitigate overfitting.

    To compare the performance of different cosmological models, we employ the AIC, defined as
        \begin{equation}
            \mathrm{AIC} = \chi^2_{\mathrm{min}} + 2 k,
        \end{equation}
    where $\chi^2_{\mathrm{min}}$ is derived from the Gaussian likelihood function $\mathcal{L}(\hat{\theta} \mid \text{data})$ evaluated at the best-fit parameters, and $k$ denotes the number of free parameters in the model. In this work, we compare the $f(T,B)$ model with the standard $\Lambda$CDM scenario using the quantity
        \begin{eqnarray}
            \Delta \mathrm{AIC} = \mathrm{AIC}_{f(T,B)} - \mathrm{AIC}_{\Lambda \mathrm{CDM}}.
        \end{eqnarray}
        
    A negative value of $\Delta \mathrm{AIC}$ indicates a preference for the $f(T,B)$ model over $\Lambda$CDM, while a positive value favors $\Lambda$CDM.

    Similarly, the BIC is calculated as
        \begin{equation}
            \mathrm{BIC} = \chi^2_{\mathrm{min}} + k \ln N,
        \end{equation}
    where $N$ denotes the number of data points used in the MCMC analysis. The BIC imposes a stricter penalty for model complexity due to the logarithmic dependence on the sample size, making it particularly sensitive to the number of parameters in large datasets.

    To quantify the relative performance of the $f(T, B)$ model against the $\Lambda$CDM benchmark, we compute the differences in these criteria
        \begin{equation}
            \Delta \mathrm{AIC} = \Delta \chi^2_{\mathrm{min}} + 2 \Delta k, \quad \Delta \mathrm{BIC} = \Delta \chi^2_{\mathrm{min}} + \Delta k \ln N,
        \end{equation}
    where $\Delta \chi^2_{\mathrm{min}}$, $\Delta k$, and $\Delta \mathrm{BIC}$ represent the differences in the minimum chi-squared values, number of parameters, and BIC values, respectively, between the two models. Smaller values of $\Delta \mathrm{AIC}$ and $\Delta \mathrm{BIC}$ indicate that the $f(T, B)$ model closely aligns with the performance of $\Lambda$CDM, suggesting competitive consistency with the observational data. 
    \begin{table}[H]
    \centering
        \def\arraystretch{1.5}
        \resizebox{0.8\textwidth}{!}{%
    \begin{tabular}{cccccccc}
    \hline \hline
        Data sets & Models & $\chi_{\rm min}^2$ & $\Delta\chi_{\rm min}^2$ & AIC & $\Delta$AIC & BIC & $\Delta$BIC \\[1ex]
    \hline 

        \raisebox{-1ex}{CC+PPS} & $f(T, B)$ & 1668.459 & \raisebox{-1ex}{$-14.023$} & 1678.459 & \raisebox{-1ex}{$-10.023$} & 1705.747 & \raisebox{-1ex}{$0.892$} \\
                         & $\Lambda$CDM & 1682.482 & & 1688.482 & & 1704.855 & \\

    \hline

        \raisebox{-1ex}{CC+PPS+DESI DR2} & $f(T, B)$ & 1673.364 & \raisebox{-1ex}{$-11.981$} & 1683.364 & \raisebox{-1ex}{$-7.981$} & 1710.678 & \raisebox{-1ex}{$2.945$} \\
                         & $\Lambda$CDM & 1685.345 & & 1691.345 & & 1707.733 & \\

    \hline \hline
    \end{tabular}}
        \caption{Goodness-of-fit statistics and comparative analysis of the $f(T, B)$ gravity model. This table presents a detailed summary of the goodness-of-fit statistics for the numerical $f(T, B)$ gravity model. It reports the minimum chi-squared value ($\chi^2_{\mathrm{min}}$), along with the corresponding differences in the AIC ($\Delta \mathrm{AIC}$) and the BIC ($\Delta \mathrm{BIC}$). These statistical indicators are directly compared to those of the standard $\Lambda$CDM model, providing a quantitative evaluation of the relative fit quality and explanatory strength of both cosmological frameworks.}
        \label{Table: statistical values}
    \end{table}
    
    In information criterion analysis, the nuisance parameter for the absolute magnitude of supernova remains constant across both models and is therefore not counted in the number of model comparison parameters. This comparative statistical analysis illuminates the strengths and limitations of the $f(T, B)$ model relative to the established $\Lambda$CDM paradigm, offering valuable insights into its viability within the broader cosmological context.

\subsection{Cosmological parameters}
    This study investigates the evolution of key cosmological diagnostics, including the effective, DE EoS parameter, the statefinder parameters, and the ${ \rm Om}(z)$ diagnostic, using constraints obtained from multiple observational datasets. Being an important characteristic that traces the dynamics of expansion of the Universe, the deceleration parameter provides us with insights into the transition from the matter-dominated decelerated period of time to accelerated expansion powered by DE, providing us with a reliable way to understand the predictive power of the model under discussion. Within the scope of the $f(T, B)$ approach, we can see how the changing narrative of our Universe is described, with the graph showing the points of acceleration initiation obtained by solving the Friedmann equations. Having been chosen with due regard for cosmological standards, the initial conditions ensure that the model agrees with observations while allowing the study of new gravitational interactions. A comprehensive formulation of the deceleration parameter, defined as $q = -\ddot{a}/aH^2$, is presented as follows
        \begin{eqnarray}
            q(z) = -1 + \frac{(1+z)H'}{H}  \, .
        \end{eqnarray}
    \begin{figure}[H]
        \centering
        \resizebox{\textwidth}{!}{%
        \includegraphics[width=8.5cm, height=7.0cm]{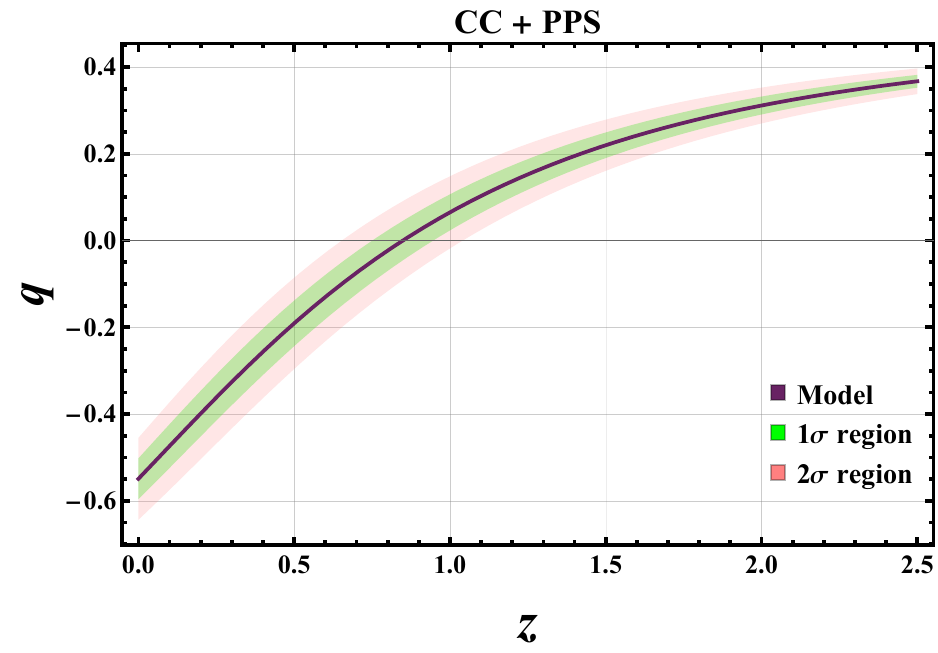}~~~~
        \includegraphics[width=8.5cm, height=7.0cm]{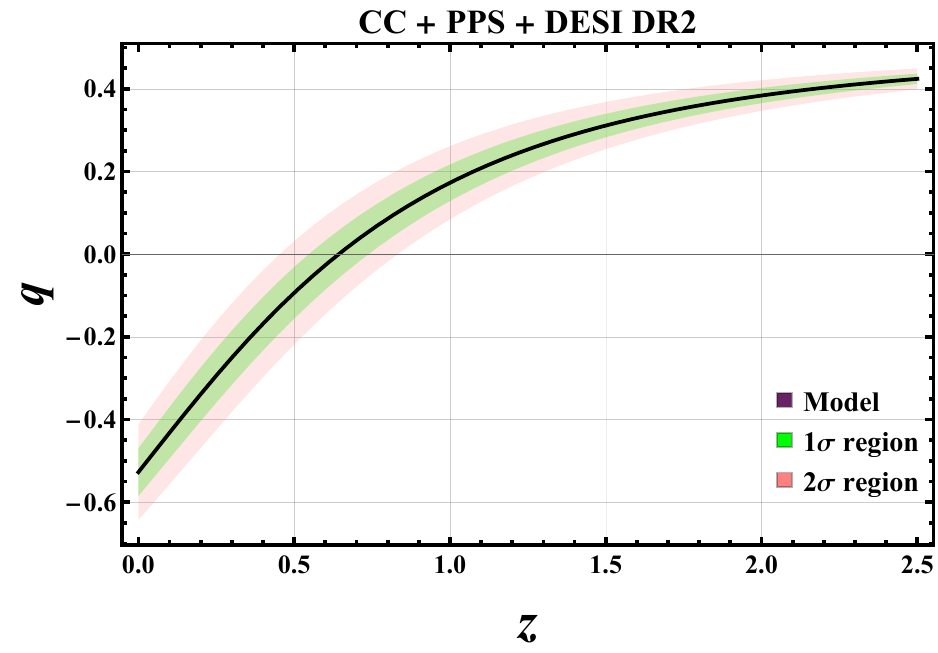}}
        \caption{Behavior of the deceleration parameter using the combined datasets.}
        \label{FIG: q}
    \end{figure}

    The graph in figure~\ref{FIG: q} represents the evolution of the deceleration parameter with the redshift $z$ for the $f(T, B)$ gravity, considering data sets from CC+PPS (left) and from CC+PPS+DESI DR2 (right). In addition, figure~\ref{FIG: q} clearly demonstrates that the constrained model parameters from the integrated data sets indicate a promising transition of the deceleration parameter $q$ from positive values, which denote deceleration in the early time, to negative values, denoting the acceleration at present times. In particular, the deceleration parameter $q_0$ at the present time is found to have values of $-0.555$ and $-0.550$ for the CC+PPS and CC+PPS+DESI DR2 data sets, respectively, consistent with the observationally expected range of $q_0 = -0.528^{+0.092}_{-0.088}$ in the latest works. The results above show a promising transition from deceleration to acceleration at the transition redshifts $z_t = 0.850$ and $z_t = 0.675$ for the CC+PPS and CC+PPS+DESI DR2 data sets, respectively, consistent with current observations. These results are consistent with the observational restrictions already well known, such as the value of $H(z)$ for approximately $z\approx 2.3$ obtained by the use of BAO, and the transition redshift $z_t = 0.60^{+0.21}_{-0.12}$ \cite{Yang_2020_2020_059}, $z_t = 0.7679^{+0.1831}_{-0.1829}$ \cite{Capozziello_2014_90_044016}, $z_t = 0.74 \pm 0.5$ \cite{Farooq_2013_766} and reported in recent studies.

\begin{figure}[H]
        \centering
        \resizebox{\textwidth}{!}{%
        \includegraphics[width=8.5cm, height=7.0cm]{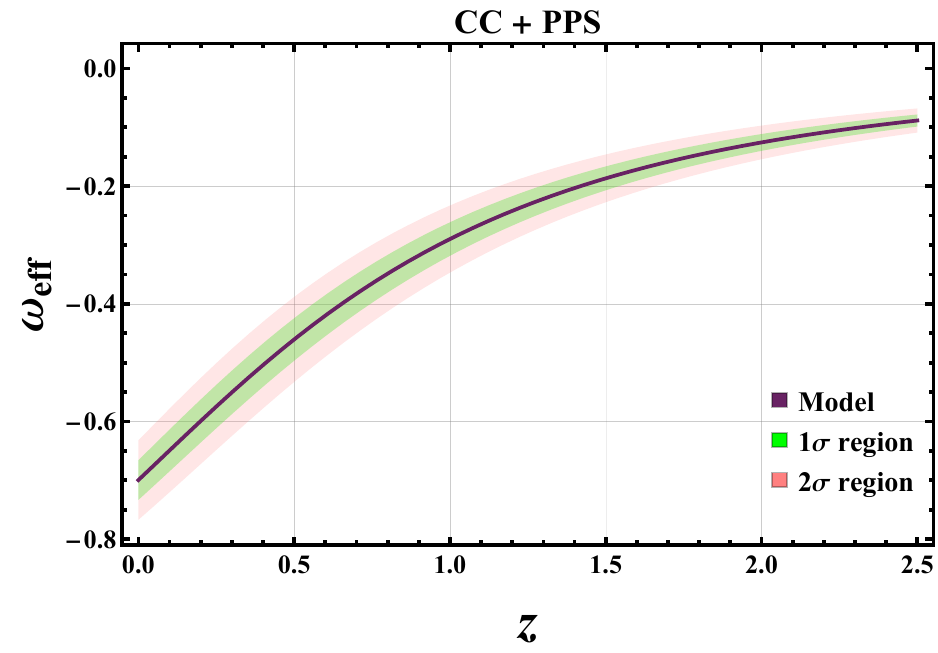}~~~~
        \includegraphics[width=8.5cm, height=7.0cm]{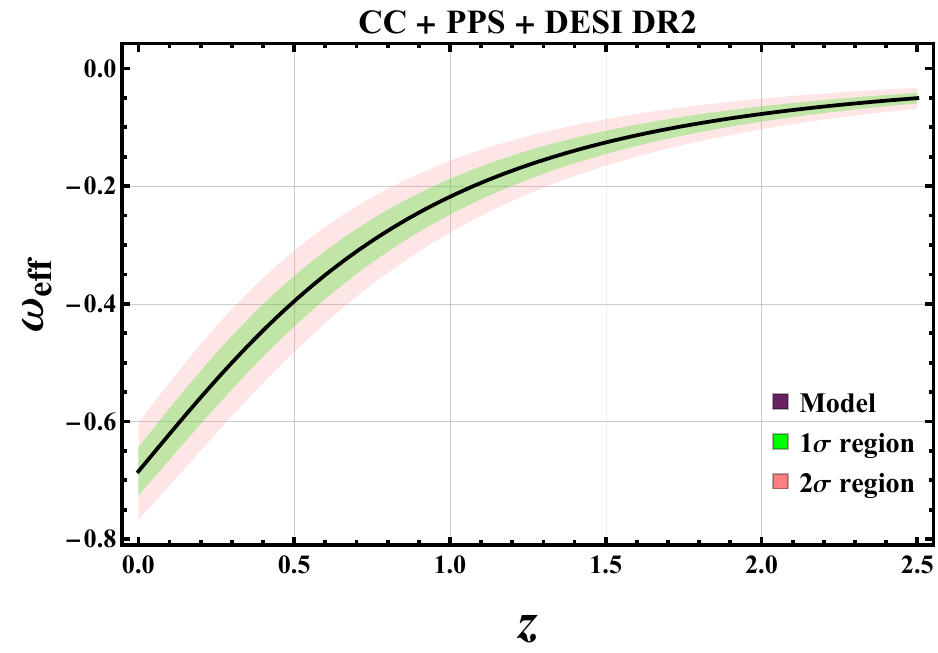}}\\
        \resizebox{\textwidth}{!}{%
        \includegraphics[width=8.5cm, height=7.0cm]{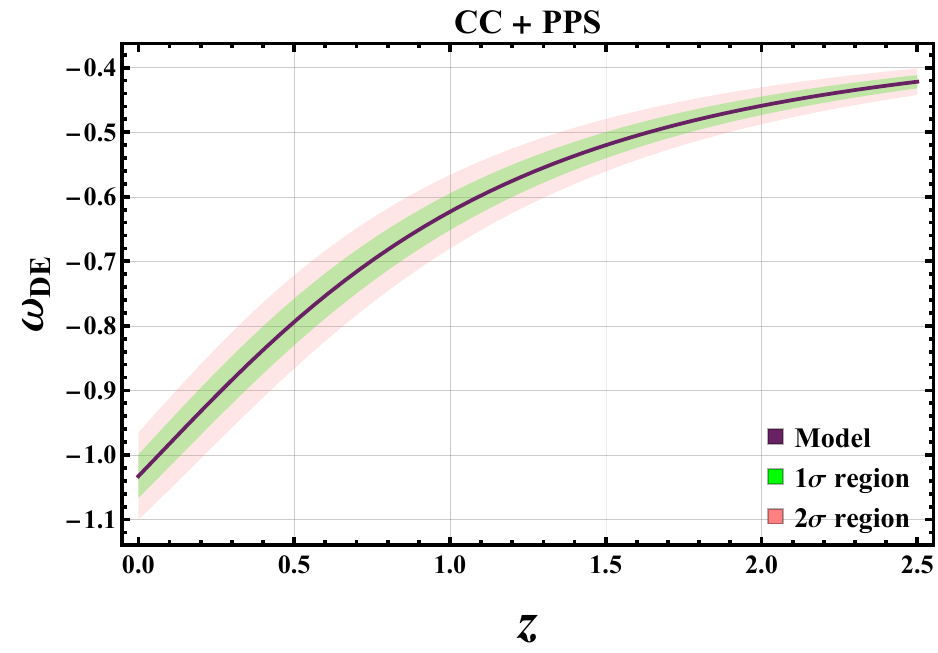}~~~~
        \includegraphics[width=8.5cm, height=7.0cm]{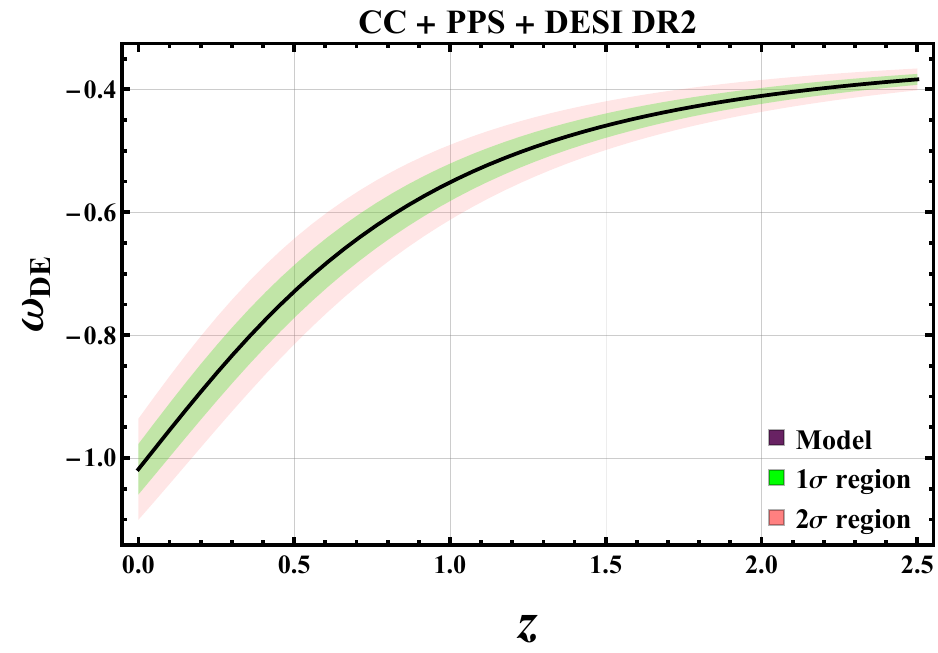}}
        \caption{Behavior of the effective and DE EoS parameter using the combined datasets.}
        \label{FIG: eos}
    \end{figure}

    The effective EoS parameter is one of the most effective tools used for identifying the nature of the energy components driving the expansion of the Universe. The behavior of $\omega_{\text{eff}}$ as a function of redshift in the $f(T, B)$ gravity theory using the CC+PPS and CC+PPS+DESI DR2 data sets is depicted in figure~\ref{FIG: eos}. The present values of $\omega_{\text{eff}}$ range between $-0.703$ and $-0.70$, consistent with properties of quintessence-type DE component. There seems to be a slight steepening towards smaller values of redshift, indicative of the dynamic nature of the DE component. Nevertheless, $\omega_{\mathrm{eff}}$ remains larger than $-1$ during the entire evolution, thereby avoiding the phantom regime.

    The variation of the effective EoS parameter, $\omega_{\text{eff}}$, versus the redshift parameter is shown in Figure~\ref{FIG: eos}, which is derived by using the corresponding DE density and pressure. In the same manner, it depicts the effect of energy sources on cosmic history via the DE EoS parameter, $\omega_{\text{DE}}$. By deriving the energy density and pressure of the DE, as shown in Figure~\ref{FIG: eos}, the result shows the change in $\omega_{\text{eff}}$ according to the redshift. The current values of the EoS parameters of DE, $\omega_{\text{DE}}(z = 0)$, are given by $-1.03289$ and $-1.00413$ for the CC+PPS and CC+PPS+DESI DR2 samples, respectively, implying the phantom behavior at low redshift and a tendency to reach around $-1.32$ at later times. These results agree with earlier cosmological works, like $\omega_{\text{DE}} = -1.035^{+0.055}_{-0.059}$, from the Supernovae Cosmology Project, Planck 2018 result of $\omega_{\text{DE}} = -1.03 \pm 0.03$, and WMAP+CMB value of $\omega_{\text{DE}} = -1.079^{+0.090}_{-0.089}$. Figure~\ref{FIG: eos} below also represents $\omega_{\text{eff}}$ in terms of redshift $z$ for the $f(T, B)$ theory of gravity. The evolution of the effective EoS parameter $(\omega_{\rm eff})$ indicates a late-time accelerating phase of the Universe. At low redshifts, $\omega_{\rm eff}$ remains sufficiently negative to satisfy the condition for accelerated expansion, $(\omega_{\rm eff}<-1/3)$. Thus, the model naturally exhibits stronger late-time acceleration, while the acceleration becomes less pronounced at higher redshifts. Similar uniformity in the behavior of $\omega_{\text{DE}}$ in both datasets confirms the power of the $f(T, B)$ theory to represent the expansion of the Universe. This change is consistent with the prediction of the theoretical model, where the higher-order curvatures are included to understand the Universe dynamics without the presence of a cosmological constant. This change is further validated by the change in $\omega_{\text{DE}}$, and makes it a promising alternative to $\Lambda$CDM.

\subsection{The statefinder diagnostic}
    The statefinder diagnostic, introduced by Sahni et al. \cite{Sahni_2003_77_201}, is an effective geometric tool for discriminating between different DE models. It uses a set of two parameters $(r, s)$; here $r$ represents the jerk parameter, which is the third derivative of the scale factor, while $s$ is a constructed parameter. These parameters are defined as
        \begin{eqnarray}
            r = \frac{\dddot{a}}{aH^3}, \quad s = \frac{r - 1}{3(q - \frac{1}{2})},
        \end{eqnarray}
    where $H$ is the Hubble parameter and $q$ is the deceleration parameter. On this diagnostic plane, the standard cosmological model based on the cosmological constant and cold dark matter, also known as $\Lambda$CDM, corresponds to a fixed point at $(r, s) = (1, 0)$. Values of $(r < 1, s > 0)$ typically indicate a quintessence-like behavior, where the EoS parameter satisfies $\omega > -1$, whereas $(r > 1, s < 0)$ points to a phantom-like regime with $\omega < -1$.

\begin{figure}[H]
        \centering
        \resizebox{\textwidth}{!}{%
        \includegraphics[width=8.5cm, height=7.0cm]{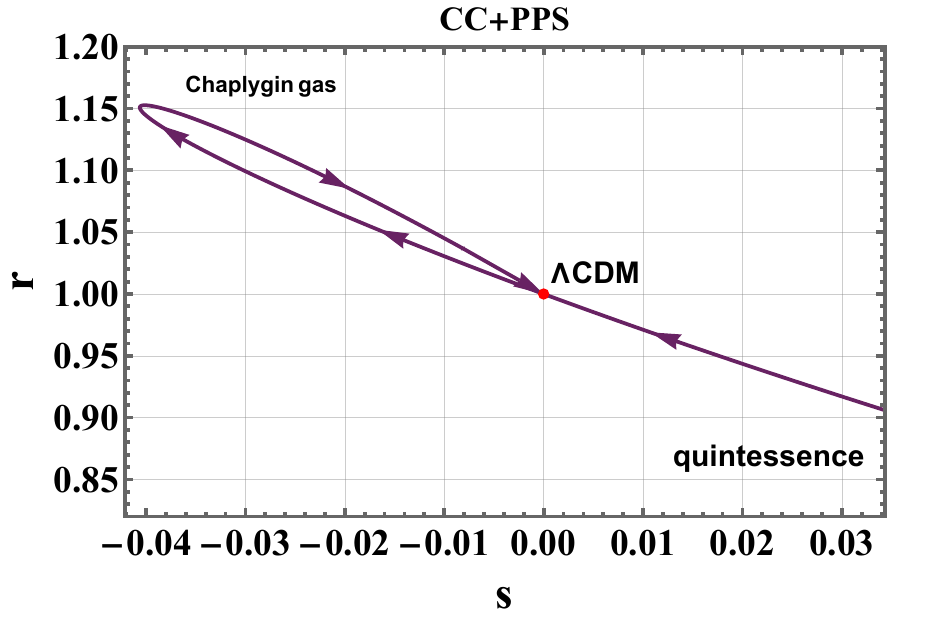}~~~~
        \includegraphics[width=8.5cm, height=7.0cm]{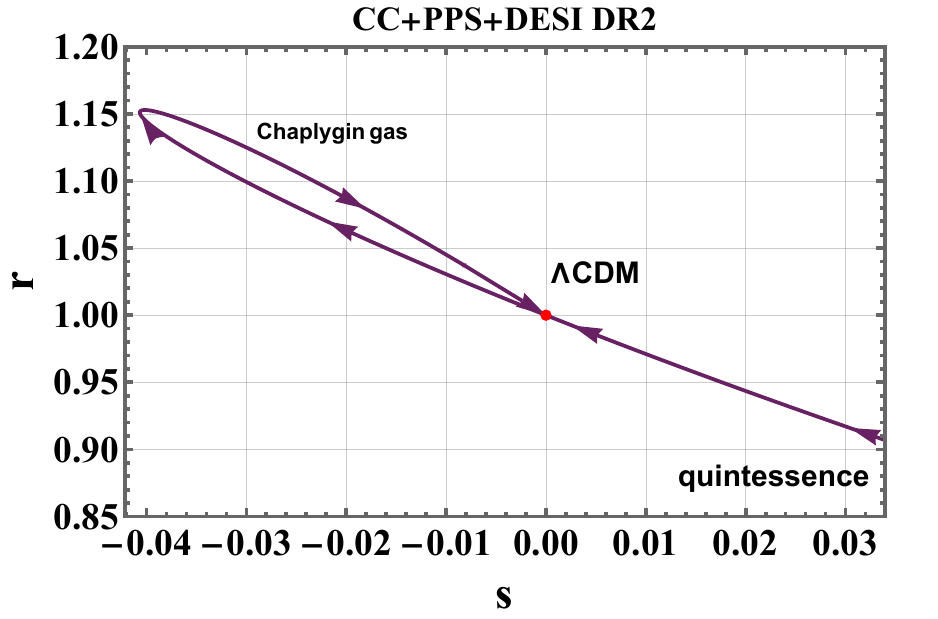}}
        \caption{Evolution trajectories of the model in the $ r $-$ s $ plane, derived using parameter values constrained by the combined CC, PPS, and DESI DR2 datasets.}
        \label{FIG: rs}
    \end{figure}

    The trajectories of the $(r, s)$ parameters in the $f(T, B)$ gravity theory are presented in figure~\ref{FIG: rs}. In contrast to the $\Lambda$CDM model, which always corresponds to a fixed point in the above plane, $f(T, B)$ gravity has an evolving path as it should be for any modified gravity theory. The line passes through the point associated with the $\Lambda$CDM model, indicating that this model reproduces the standard evolution in certain periods. Even though the statefinder trajectory traces the quintessence-like and Chaplygin-gas-like parameter spaces during cosmological history, the state of the Universe is situated very close to the $\Lambda$CDM attractor, $(r,s)=(1,0)$. As a result, the model operates like the $\Lambda$CDM model in the current epoch. At the same time, the trajectory passing through the Chaplygin-gas-like space indicates the possibility of a dark sector unification, when the effective gravitational dynamics is responsible for both dark matter and DE. Despite the fact that the effective EoS parameter is always larger than minus one, indicating the quintessence-like behavior, the $f(T, B)$ gravity may have a phantom-like expansion as well. This property is quite typical of modified gravity theories including $k$-essence models \cite{Picon_2001_63_103510} where phantom-like evolution may occur even when the parameter $\omega_{\rm eff}$ is larger than minus one \cite{Cai_2016_79}. Statefinder diagnostics clearly illustrate the flexibility of the $f(T, B)$ theory to explain a wide range of cosmological scenarios. The model undergoes changes across different eras of cosmic dynamics smoothly; therefore, it provides a unified framework for describing the history of cosmic evolution. Paths followed by the system in the $r-s$ parameter space give an exclusive geometrical signature that clearly separates the $f(T,B)$ scenario from the static $\Lambda$CDM fixed point $(1,0)$. One of the unique properties of the evolutionary paths is the crossing of the Chaplygin gas regime ($r > 1, s < 0$) by the model prior to its convergence toward the de Sitter attractor. This kind of excursion implies that during the phase-transition era, the torsion-boundary term plays the role of the unified dark sector component, such as a Chaplygin fluid. This looping behavior offers a distinctive imprint: the $f(T, B)$ model is able to describe the $\Lambda$CDM-like expansion at certain times but not the dynamics in between them.
    \begin{figure}[H]
        \centering
        \resizebox{\textwidth}{!}{%
        \includegraphics[width=8.5cm, height=7.0cm]{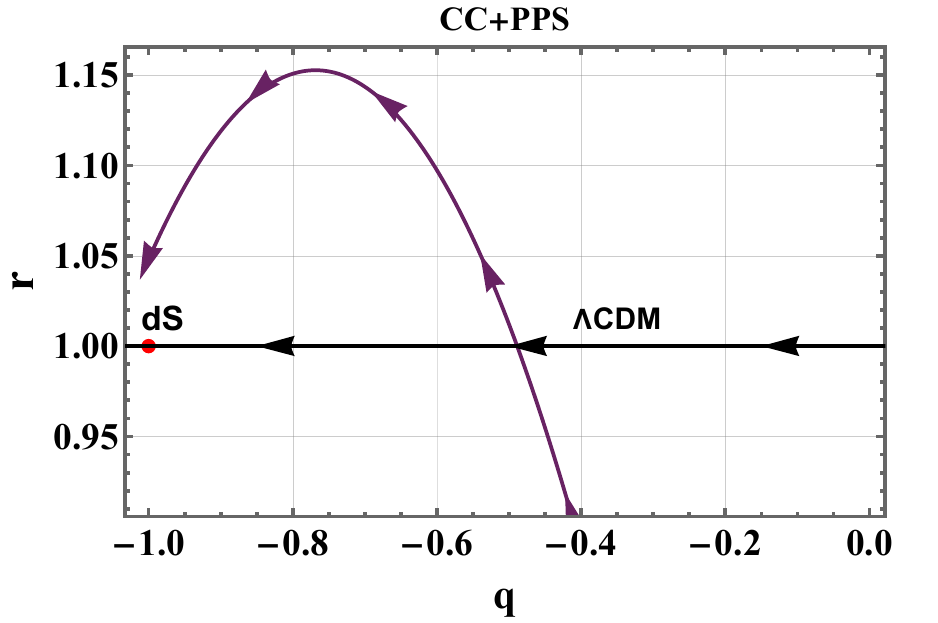}~~~~
        \includegraphics[width=8.5cm, height=7.0cm]{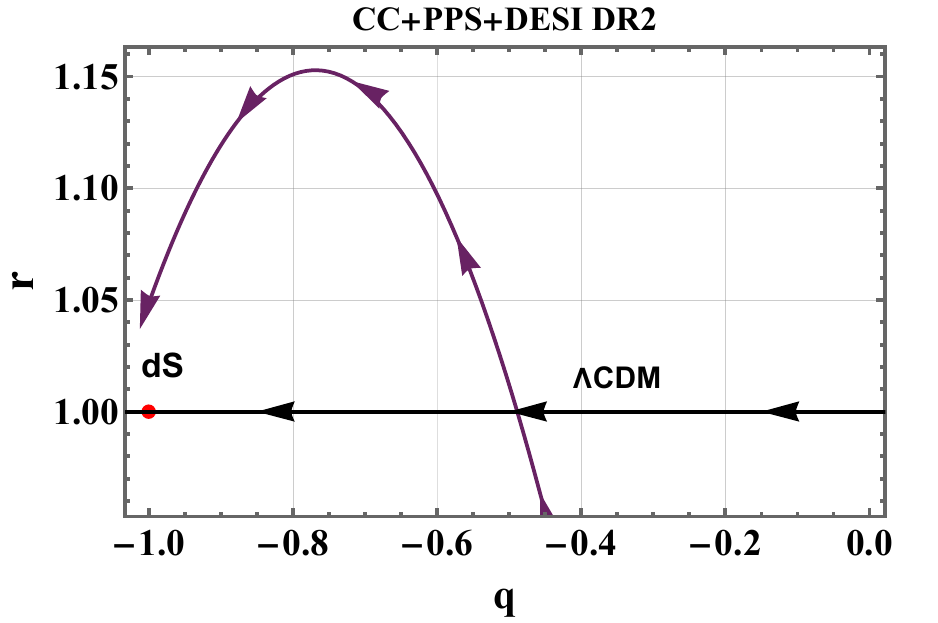}}
        \caption{Evolution trajectories of the model in the $ r $-$ q $ plane, derived using parameter values constrained by the combined CC, PPS, and DESI DR2 datasets.}
        \label{FIG: rq}
    \end{figure}

    Figure~\ref{FIG: rq} displays the dynamical evolution of the cosmological model in the $r-q$ parameter space, where $q$ is the deceleration parameter while $r$ stands for the jerk parameter. Over time, the model's evolution trajectory approaches the de Sitter (dS) point at the coordinates $(q, r) = (-1, 1)$, implying accelerated expansion due to the presence of the cosmological constant-like matter component. In the asymptotic future, the model's evolution trajectory converges to the statefinder pair $(r, s) = (1, 0)$ or $(q, r) = (-1, 1)$ and thus agrees very well with the $\Lambda$CDM model or de Sitter Universe. Such evolutionary scenario allows us to understand how the model smoothly passes from the matter-dominated epoch to the one which is characterized by domination of DE. In the $r-q$ plane, the model's trajectory shows a smooth but non-trivial transition from the conventional matter-dominated epoch $(q = 0.5, r = 1)$ to the accelerated regime of expansion. The curved trajectory towards the de Sitter (dS) point $(-1, 1)$ demonstrates that the start of the cosmic acceleration within the $f(T,B)$ gravity is much more complicated than in the standard GR framework.

\subsection{The \texorpdfstring{$\text{Om}(z)$}{} diagnostic}
    The $\text{Om}(z)$ diagnostic serves as a pivotal theoretical framework for differentiating alternative cosmological models from the standard $\Lambda$CDM paradigm. This diagnostic method enables researchers to probe the properties of the DE component by analyzing the redshift-dependent behavior of the expansion of the Universe. Specifically, the $\text{Om}(z)$ function is mathematically expressed as
        \begin{eqnarray}
            \text{Om}(z) = \frac{\left(\frac{H(z)}{H_0}\right)^2 - 1}{(1 + z)^3 - 1}
        \end{eqnarray}
    
    This formulation leverages the Hubble parameter $H(z)$ relative to its present-day value $H_0$, offering insights into the evolutionary dynamics of DE and facilitating the identification of deviations from the $\Lambda$CDM model.

    The two-point difference diagnostic, defined as
        \begin{equation}
            \text{Om}(z_1, z_2) = \text{Om}(z_1) - \text{Om}(z_2),
        \end{equation}
    where $z_1 < z_2$, provides a robust tool for characterizing DE dynamics. A positive value of $\text{Om}(z_1, z_2)$ suggests a quintessence-like behavior ($\omega > -1$), while a negative value indicates a phantom-like phase ($\omega < -1$). Within the $\Lambda$CDM framework, the $\text{Om}(z)$ diagnostic serves as a null test, as established by Sahni et al. \cite{Sahni_2008_78_103502}. Its sensitivity to the EoS parameter has been further validated through subsequent studies, including Qi et al. \cite{Qi_2018_18_066}, Zheng et al. \cite{Zheng_2016_825_17}, Lohakare et al. \cite{Lohakare_2024_MNRAS}, and Ding et al. \cite{Ding_2015_803_L22}. A constant $\text{Om}(z)$ across redshift implies that the DE behaves as a cosmological constant. The slope of $\text{Om}(z)$ further distinguishes DE models: a positive slope corresponds to a phantom phase ($\omega < -1$), while a negative slope aligns with quintessence ($\omega > -1$). 

\begin{figure}[H]
        \centering
        \resizebox{\textwidth}{!}{%
        \includegraphics[width=8.5cm, height=7.0cm]{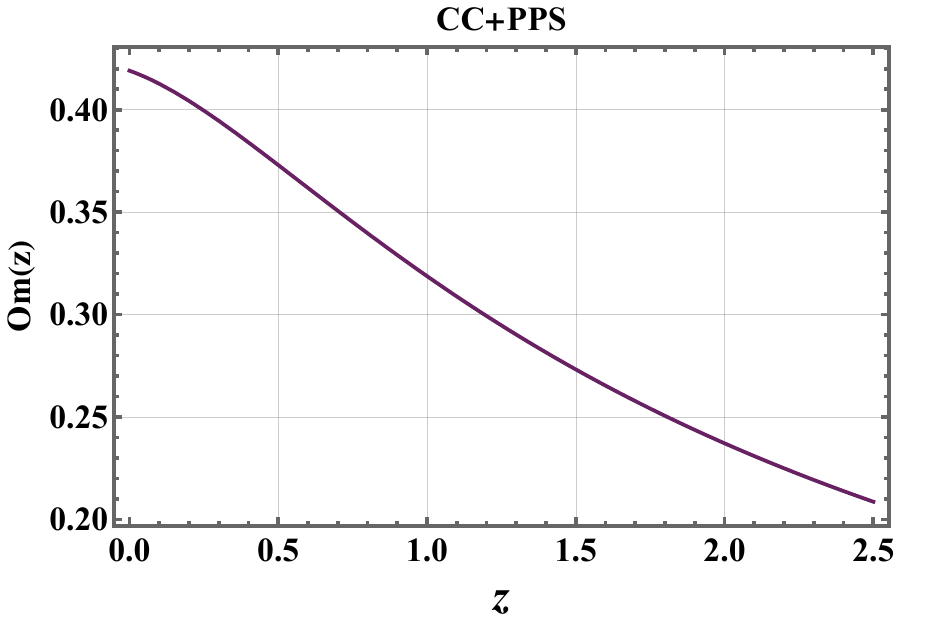}~~~~
        \includegraphics[width=8.5cm, height=7.0cm]{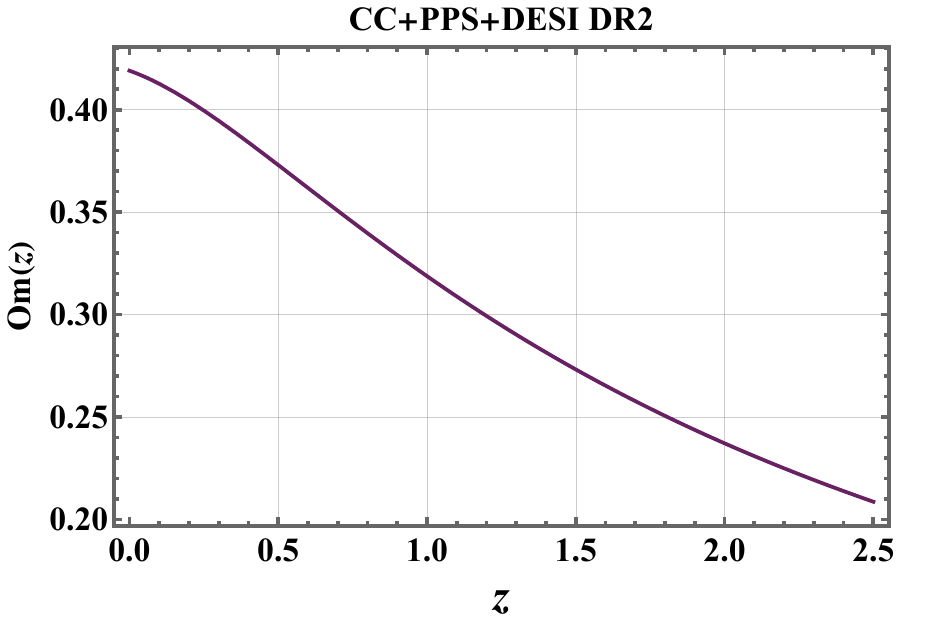}}
        \caption{Profile of the $\text{Om}(z)$ diagnostic parameter for the cosmological model, derived using constrained coefficients from combined datasets.}
        \label{FIG: Om}
    \end{figure}

    Figure~\ref{FIG: Om} displays the reconstructed $\text{Om}(z)$ parameter, obtained using the best-fitting data, as a function of redshift. As it can be seen from the figure, there is a clear trend of decreasing $\text{Om}(z)$ with respect to the increase in redshift. Such a specific trend, which manifests itself by a particular inclination of the ${\rm Om}(z)$ curve, is a perfect indicator of deviation from the cosmological constant scenario. The curves continue to lie above the current matter-density parameter $\Omega_{\rm m0}\simeq 0.32$ for small values of $z$, but they drop down as $z>1$. This means that the torsion-boundary coupling leads to an effective dynamical gravity sector which becomes important at relatively late epochs. This particular feature of the geometry originates from the presence of the non-linear coupling parameters $\alpha$ and $\beta$, which allow to have an evolving DE component. Such behaviour serves as a good observational signature: the amount of the deviation from the $\Lambda$CDM flat line (${ \rm Om}(z)=$const.) could serve as an indicator to distinguish between the gravity and DE scenarios. Similar behaviors were observed in $\text{Om}(z)$ for both data samples considered here. In particular, the ${ \rm Om}(z)$ diagnostic decreases at higher redshifts ($z > 1$), which is an indication of deviation from the $\Lambda$CDM model within the $f(T, B)$ gravity scenario.
    The discrepancy is due to the effects of extra torsional and boundary-term effects in the $f(T,B)$ theory, which affect the cosmic evolution process differently from the case of the standard $\Lambda$CDM theory. Furthermore, the form of the ${ \rm Om}(z)$ diagnostic plot shows that the theory can explain the accelerated expansion of the cosmos without resorting to a cosmological constant, but rather via a modified-gravity approach.

\section{Conclusion} 
\label{sec6:Conclusion}
    In this work, we examine the theoretical framework of modified teleparallel gravity with the inclusion of the boundary term in the action and investigate its cosmological implications by considering the power-law model $f(T, B) = -T + \alpha(-B)^\beta$, with the aim of addressing late-time accelerated expansion and the DE problem. In this framework, $T$ denotes the torsion scalar and $B$ represents the boundary term, whose presence allows for departures from standard teleparallel dynamics and provides a unified description that connects torsion and curvature-based formulations, reproducing $f(T)$ and $f(R)$ gravity in appropriate limits. The viability of the model is assessed by confronting its theoretical predictions with observational data, constraining the cosmological and model parameters through MCMC analysis using late-time datasets, namely CC, the PPS, and the DESI BAO Data Release 2 (DR2), and comparing its performance with the standard $\Lambda$CDM model.
    
    By taking into account different combinations of observational datasets, we obtain the constraints $H_{0}=70.132\pm 1.151~{\rm km~s^{-1}~Mpc^{-1}}$ from the CC+PPS combination and $H_{0}=69.120\pm 1.032~{\rm km~s^{-1}~Mpc^{-1}}$ from the combined CC+PPS+DESI DR2, yielding values that lie between the early-Universe estimate inferred from Planck and the local SH0ES measurement, indicating that the model remains consistent with current late-time cosmological observations. The kinematic analysis also indicates a smooth transition from decelerated expansion to late-time acceleration, with the transition redshift in the range $z_{t}\simeq 0.850,\, 0.675$ for CC+PPS and CC+PPS+DESI DR2, respectively. The corresponding present value of the deceleration parameter, $q_{0}\simeq -0.55$, is consistent with the expected behavior of an accelerating Universe.

    An important feature emerging from the analysis is the crossing of the phantom divide in the DE EoS. At low redshift, $\omega_{\rm DE}$ becomes slightly smaller than $-1$ and approaches approximately $\omega_{\rm DE}\simeq -1.32$ at late times. This behavior points to an evolving effective DE sector rather than a strictly constant cosmological constant. Such a trend is broadly in line with the indications reported in the DESI 2024 and DESI DR2 analyses, where the data show a preference, at the level of $2.8\sigma$ to $4.2\sigma$, for evolving DE with $\omega_{0}>-1$ and $\omega_{a}<0$ over a rigid cosmological constant. At the same time, these indications should be interpreted with care, since possible systematics in SNIa and BAO datasets, as well as potential departures from the distance-duality relation, may affect the inferred evidence for DE evolution. Within this context, the present $f(T,B)$ model provides a geometric mechanism through which such an effective evolution can arise.

    Our results, for the obtained constraints $\beta$ from the model parameter, indicate that, within the adopted parameterization and prior ranges, the observational data favor a non-zero correction to the boundary term. The model comparison provides a useful, though not entirely one-sided, indication of the statistical performance of the $f(T,B)$ scenario. The BIC mildly favours $\Lambda$CDM, mainly because the concordance model contains fewer free parameters, with $\Delta{\rm BIC}\simeq 0.892,\, 2.945$. In contrast, the AIC favours the $f(T,B)$ model, giving $\Delta{\rm AIC}\simeq -10.023,\, -7.981$. This difference between the two criteria is not unexpected: the AIC is more sensitive to improvements in fit, whereas the BIC penalizes the additional parameters more strongly. Thus, the data allow the extra freedom introduced by the torsion--boundary sector, but the level of statistical preference depends on the criterion used for model selection.

    The statefinder diagnostic provides another way to see how the model departs from the concordance picture. In the $(r,s)$ plane, the evolutionary trajectories do not stay at the $\Lambda$CDM fixed point $(1,0)$. Instead, they move through regions that are usually associated with Chaplygin gas-like behavior. This indicates that, although the model can closely follow the standard expansion history, its higher-order kinematic evolution is not identical to that of $\Lambda$CDM. In this sense, the statefinder analysis helps to separate the present $f(T,B)$ model from a simple reproduction of the concordance scenario. Further, this result is compatible with recent discussions of power-law $f(T,B)$ models, where such models have been shown to admit viable background evolution while allowing departures from the standard cosmological trajectory.

    A future extension of this work is to investigate the impact of the $f(T,B)$ framework on the evolution of matter density perturbations and its implications for the $S_8$ tension. This will be followed by an analysis of the stability of these perturbations and the conditions for the absence of ghost degrees of freedom, which are essential for establishing the theoretical consistency of the model. Furthermore, constraints from the early Universe, such as big bang nucleosynthesis and inflationary dynamics, would provide complementary insights and help assess the viability of the torsion--boundary framework across different stages of cosmic evolution.
    
    Although the modified gravity framework considered here does not fully resolve the existing cosmological discrepancies, it nevertheless provides an improved fit to the observational data compared to the standard $\Lambda$CDM model. This suggests that deviations from the conventional picture, whether through modifications of gravity or dynamical DE, may still carry essential clues about the underlying physics governing the late-time Universe.  At the same time, the present results should be interpreted with some caution, as they point to the need for further investigation. In particular, forthcoming high-precision data will be essential to test the internal consistency of the model and to assess its robustness across different observational probes. A more systematic statistical analysis may also be required to resolve potential degeneracies and to clarify whether the observed tensions arise from limitations of the current models or from deeper, as yet unresolved, aspects of the gravitational interaction itself.

\section*{Acknowledgment}
    L.P. acknowledges the financial support provided by the Estonian Research Council “grant PSG910 Theoretical frameworks for numerical modified gravity'' and the COSMOVERSE COST Action CA21136 \,“Addressing observational tensions in cosmology with systematics and fundamental physics''. The author S. K. Maurya acknowledges that the Ministry of Higher Education, Research, and Innovation (MoHERI) supported this research work through the project BFP/GRG/CBS/24/035.  Also, S. K. Maurya appreciates the administration of the University of Nizwa, Oman for their unwavering support and encouragement. This work was financially supported by Vellore Institute of Technology (VIT), Vellore under the Faculty Seed Grant (RGEMS) (Sanction Order No.: SG20260368). The authors would like to thank the anonymous reviewer for their valuable comments and suggestions, which helped improve the quality and clarity of the manuscript.
    

\providecommand{\href}[2]{#2}\begingroup\raggedright\endgroup

\end{document}